\documentclass[12pt,preprint]{aastex}

\newcommand{\hi}{H{\sc i}\,\,}
\newcommand{\kms}{km~s$^{-1}$}
\newcommand{\msol}{M$_{\odot}$}

\shorttitle{\hi in early-type dwarf galaxies}
\shortauthors{Bouchard, Jerjen, Da Costa \& Ott}

\begin{document}

\title{Detection of neutral hydrogen in early-type dwarf galaxies of the Sculptor
Group}

\author{Antoine Bouchard}
\affil{Research School of Astronomy \& Astrophysics, Mount Stromlo Observatory, Cotter Road, Weston Creek, ACT 2611 Australia
\\and\\
Australia Telescope National Facility, PO Box 76, Epping, NSW 1710, Australia}
\email{bouchard@mso.anu.edu.au}

\author{Helmut Jerjen, Gary S. Da Costa}
\affil{Research School of Astronomy \& Astrophysics, Mount Stromlo Observatory, Cotter Road, Weston Creek, ACT 2611 Australia}
\email{jerjen@mso.anu.edu.au, gdc@mso.anu.edu.au}

\and

\author{J\"urgen Ott}
\affil{Australia Telescope National Facility, PO Box 76, Epping, NSW 1710, Australia}
\email{Juergen.Ott@atnf.csiro.au}

\begin{abstract}
\footnotetext[1]{The Parkes telescope and the Australia Telescope Compact Array
are part of the Australia Telescope which is funded by the Commonwealth of
Australia for operation as a National Facility managed by CSIRO.} We present our
results of deep 21\,cm neutral hydrogen (H{\sc i}) line observations of five early and mixed-type
dwarf galaxies in the nearby Sculptor group using the ATNF 64m Parkes\footnotemark[1]
Radio Telescope. Four of these objects,  ESO294-G010, ESO410-G005, ESO540-G030,
and ESO540-G032, were detected in \hi with neutral hydrogen masses in the range
of $2-9\times10^5$ \msol{} ($M_{\rm HI}/L_{\rm B}$ = 0.08, 0.13, 0.16, and 0.18
$M_{\odot}/L_{\odot}$, respectively). These \hi masses are consistent with the
gas mass expected from stellar outflows over a large period of time. Higher
spatial
resolution \hi data from the Australia Telescope Compact Array\footnotemark[1]
interferometer
were further analysed to measure more accurate positions and the distribution of
the \hi gas.  In the cases of dwarfs ESO294-G010 and ESO540-G030, we find
significant offsets of 290\,pc and 460\,pc, respectively, between the position
of the \hi peak flux and the center of the stellar component. These offsets are
likely to have internal cause such as the winds from star-forming regions. The
fifth object, the spatially isolated dwarf elliptical Scl-dE1, remains
undetected at our $3\sigma$ limit of 22.5 mJy \kms{} and thus must contain less
than $10^5$ \msol{} of neutral hydrogen. This leaves Scl-dE1 as the only
Sculptor group galaxy known where no interstellar medium has been found to date.
The object joins a list of similar systems including the Local Group dwarfs
Tucana and Cetus  that do not fit into the global picture of the
morphology-density relation where gas-rich dwarf irregulars are in relative
isolation and gas-deficient dwarf ellipticals are satellites of more luminous
galaxies. 
\end{abstract}


\section{Introduction}

Dwarf galaxies are at the center of a long debated galaxy evolution puzzle.
Detailed studies of Local Group dwarfs have found increasing evidence that the
specific morphological type of any particular dwarf is strongly correlated with
the density of its local environment \citep[i.e. the morphology-density realtion][]{einasto1974,vandenbergh1994a}:
late-type dwarfs (mainly the dwarf irregulars, dIrr) tend to be further away
from more massive and more luminous stellar systems than dwarf elliptical
galaxies (dEs, including dwarf S0s, dS0, and dwarf spheroidals, dSph).  These
two main dwarf types differ in many other ways.  In general, early-types, with
their smooth ellipsoidal stellar distribution, lack an interstellar medium
(ISM), possess low angular momentum \citep[although see][ for a discussion on the
luminous dEs]{geha2003}, and have low current star formation rates
\citep[e.g.][]{mateo1998}. In contrast, the late-types exhibit an irregular
optical appearance that is dominated by H{\sc ii} regions and ongoing star
formation, large neutral hydrogen mass to luminosity ratios
\citep[e.g.][]{koribalski2004}, and higher angular momentum.

Morphologically, there is no clear boundary between the two main dwarf types,
dIrr and dE. Instead, there exist mixed-type dwarfs (dIrr/dE, e.g. Phoenix)
which exhibit properties inherent to both categories.  Furthermore, detailed
stellar population studies of Local Group dwarfs have revealed that dEs do not
have simple single-burst, old stellar populations like globular clusters, but
instead show a large variety of star formation histories (e.g.
\citealt{grebel2001}; \citealt{hensler2004}).

The morphology-density relation and the continuous range in stellar/gas
properties across dwarf galaxy families naturally suggests that evolution from
one main morphological class to the other might be an ongoing process that is
still observable in the local Universe.  Environmentally driven physical
processes such as ram pressure stripping \citep{einasto1974} and/or tidal
effects \citep{moore1996}, or internal processes like stellar winds and
supernovae explosions combined with the shallow gravitational potential of
the galaxy could be responsible for a fast morphological transformation. In such
a scenario, the mixed type dwarfs are thought to be in the transition phase from
dIrr to a dE stage.  While theoretically supported by numerical simulations
\citep{mayer2001}, \citet{grebel2003} argued against this evolutionary scenario
with two major objections: firstly, the fading of a dIrr to the typical
luminosity of a dE would produce an object with too low metallicity than what is
observed and, secondly, the removal of the ISM from a dIrr would not produce a
non-rotating dE \citep[see also][]{read2005}. However, one should note that the
absence of angular momentum is no longer considered as a defining element of the
dE class since rotation was measured in some of the bright dEs
\citep[e.g.][]{vanzee2004,derijcke2004,derijcke2003,pedraz2002,simien2002}. 

The limitation of the morphological classification is that it merely describes
the stellar distribution of a galaxy and conveys little physical information on
the objects. This classification scheme gives, at best, an imprecise view of the
stellar population (i.e. it can not distinguish between a 1 Gyr and 15 Gyr old
population) and conveys no information on the total mass, internal kinematics,
or chemical composition.
Yet, studies of Local Group dEs helped to establish
empirical relations between morphology and physical parameters such as the
morphology-density relation, but have failed to establish a direct quantitative
correlation based on solid physical principles. Consequently, it is difficult to
understand the relation between individual galaxy types and their respective
evolutionary state. In fact, the properties of dwarf galaxies in the Local Group
paint a rather complex picture.  For example, the spatially isolated, mixed
morphology galaxy LGS3 is dominated by an old stellar population but has had
sustained star formation for most of its life (\citealt{aparicio1997};
\citealt{miller2001}). It also contains $\sim$6$\times$10$^5$ \msol{} of \hi
centered on the optical component of the galaxy \citep{young1997}.  The slightly
less isolated Phoenix dwarf, also of mixed morphology, has a $\sim 10^5$ \msol{}
\hi cloud that is offset by 650\,pc (a third of the tidal radii,
\citealt{martinez1999}) from the optical center \citep{st-germain1999} while
also having a predominantly old stellar population but sustained, long term star
formation \citep{holtzman2000}.  The similarly isolated Tucana dwarf has had no
extended star formation \citep[e.g.][]{dacosta1998} and contains no \hi within
its optical boundaries.  There is, however, a hydrogen cloud projected on the sky
near the galaxy, which might be related \citep[see][]{oosterloo1996}.  Further,
the Local Group dSph Sculptor apparently possesses two \hi clouds distributed
symmetrically on either side of the optical center (\citealt{knapp1978};
\citealt{carignan1998}; \citealt{bouchard2003}).  The central regions (within
$\sim1.7$ core radii or 230\,pc) are, however, gas free and the galaxy itself
contains only very old stars ($>$ 10\,Gyr) with no evidence of any ongoing star
formation (\citealt{hurley1999}; \citealt{monkiewicz1999}).

Due to this rather complex picture drawn from Local Group dwarfs, one needs to
study a larger sample of galaxies --- therefore beyond the Local Group --- to
establish a quantitative measure of the environmental influence on a dwarf
system and to put it into a relation with the fundamental properties of the
galaxies (e.g.  stellar population, etc.).  Such a relation needs to take into
account the quantity of ISM and its distribution. 
\citet[][ and references therein]{conselice2003} previously reported that $\sim
15\%$ of early-type dwarfs in the dense Virgo cluster have $M_{\rm HI} \gtrsim
10^7 M_{\odot}$. These galaxies are potentially newly accreted members or on
orbits that never lead them in the cluster's center. In addition, they have \hi
properties similar to Local Group transition and late-type dwarfs; the Local
Group early-type
dwarfs have much lower \hi content. It is argued that the Virgo cluster acts as
an "evolutionary change engine" and that these \hi rich dEs were dIrrs that have
recently made the transition by means of ram pressure or Kelvin-Helmholtz
instabilities (i.e. the morphology-density relation at work). Alternatively,
\citet{buyle2005} suggests that an enhanced star formation rate due to ram
pressure and gravitational interactions, will accelerate the gas depletion in
such systems.

Both these studies have targeted relatively bright dEs ($L_{\rm B} \gtrsim 10^7
L_{\odot}$) in a dense environment, somewhat different to the situation of Local
Group
dEs. In order to assess the overall validity of the morphology-density relation
and the relative importance of the environment in dwarf galaxy evolution, we
must now study a regime of low mass galaxies in a low density environment. Under
such conditions, the environmental impact on galaxy evolution should be
minimized while the shallow potential well of the low mass dEs will help to amplify the
effect of internal mechanisms (e.g. stellar winds, supernova explosions).

\begin{figure}
\begin{center}
\plotone{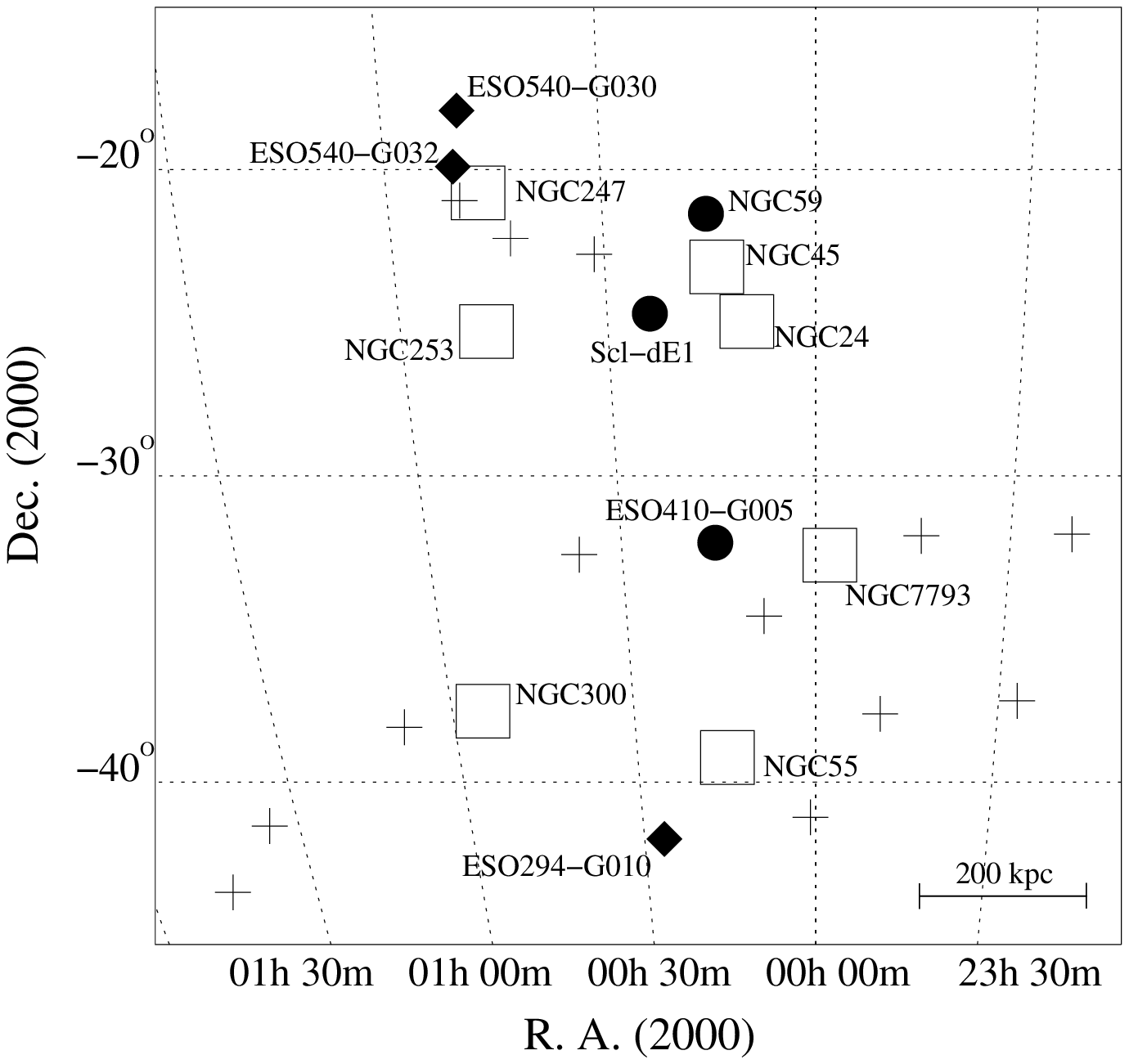}
\caption{Sky distribution of all known members of the Sculptor group. Squares
show the main members of the group, the filled circles denote the early-type
dwarfs, the filled diamonds are for mixed types, and +'s are for late-type
dwarfs. The scale of the Figure (at a distance of 2 Mpc) is indicated in the 
lower right corner.}\label{distribution} \end{center}
\end{figure}

The Sculptor group (Fig.~\ref{distribution}), with its sparse galaxy
distribution, provides an ideal laboratory to study galaxies in such an environment
without having to hunt down truly isolated objects. Due to its proximity to the
Local Group (1.5 to 4\,Mpc, \citealt{jerjen1998}; \citealt{karachentsev2003}),
detailed studies of stellar populations
\citep{karachentsev2000,skillman2003a,olsen2004}, ISM contents
\citep{cote1997,deblok2002}, and chemical compositions \citep{skillman2003b} of
Sculptor galaxies are available. 

There are six early-type dwarf galaxies known in the Sculptor group and prior to
this study five have not been detected in H{\sc i}. The sixth, NGC 59, is a dS0
galaxy for which \citet{cote1997} reported detections in both \hi
(5.4$\times10^6$ \msol{}) and H$\alpha$.  The five other galaxies are: the
dS0/dIrr ESO294-G010, the two dE/dIrr ESO540-G030 and ESO540-G032, the dSph
ESO410-G005, and the dE Scl-dE1.  For all but Scl-dE1, stellar photometry based
on Hubble Space Telescope data reveals statistical evidence for the presence of
blue stars in the central regions of these galaxies
\citep{karachentsev2000,jerjen2001,karachentsev2002,karachentsev2003} although
the possibility that such stars result from photometric errors in the crowded
central fields cannot be excluded at this time. In addition to these possibly
young stars, \citet{jerjen1998} have identified a small H{\sc ii} region in
ESO294-G010.

This paper presents the results of a deep \hi study of these five early-type
Sculptor group dwarfs and addresses the central question as to whether the light
distribution of a galaxy (i.e. morphology) is sufficient to predict
ISM content. The paper is organised as follows: the observations are described
in section 2, section 3 describes the \hi properties of our sample galaxies, and
section 4 provides a discussion of the Sculptor group. The main conclusions can
be found in section 5. 

\section{Observations}

\subsection{Sample properties}
We have observed the five lowest luminosity objects of the Sculptor group. They
are also the only five objects of the group where no \hi detection were
previously reported.  General properties of these five dwarfs are described in
detail in \citet{jerjen1998,jerjen2000}. In Table \ref{objects}, we list some of
their physical parameters.  The first two columns provide the coordinates. The
distances $D$ in Mpc have previously been measured using the tip of the red
giant branch technique
\citep{karachentsev2000,karachentsev2002,karachentsev2003}. The apparent
$B$-magnitudes $m_{\rm B}$ and extinction are given in \citet{jerjen2000}.  This
allowed us to calculate the luminosities, $L_{\rm B}$ (we adopted a value of
M$_{\odot,B}$ = 5.5, \citealt{bessell1998}), and absolute magnitudes, M$_B$.
For ESO294-G010, the optical, spectroscopic velocity \citep{jerjen1998} is also listed in Table
\ref{objects}.

\subsection{\hi observations}

In 2004 October, the 64\,m ATNF Parkes Radio Telescope was employed to obtain
deep, high spectral resolution \hi spectra toward the five Sculptor dwarf galaxies.
We used the Multibeam instrument in MX (beam-switching) mode along with the narrowband
correlator in the MB7\_8\_1024 configuration provide 8 MHz bandwidth divided
into 1024 channels. The observations were done at a central frequency of 1418
MHz, resulting in an \hi velocity coverage from $-300$ to $1300$ \kms{} with
channel widths of 1.6 \kms{} and a beam size of 14.1$^{\prime}$. This
configuration enabled us to observe with seven beams simultaneously, keeping one
beam on-source with the six others on adjacent sky, alternating the on source
beam every two minutes.
The integration times for
the target galaxies were:
294\,min for ESO294-G010, 
210\,min for ESO410-G005, 
336\,min for ESO540-G030, 
and 224\,min for ESO540-G032. These resulted in clear detections in H{\sc i}.
The RMS values are listed in Table \ref{detections}.  In the case of Scl-dE1,
434\,min of on-source integration yielded a spectral RMS of 1.2\,mJy but no
detection.

\begin{figure}
\begin{center}
\includegraphics[angle=270, width=\textwidth]{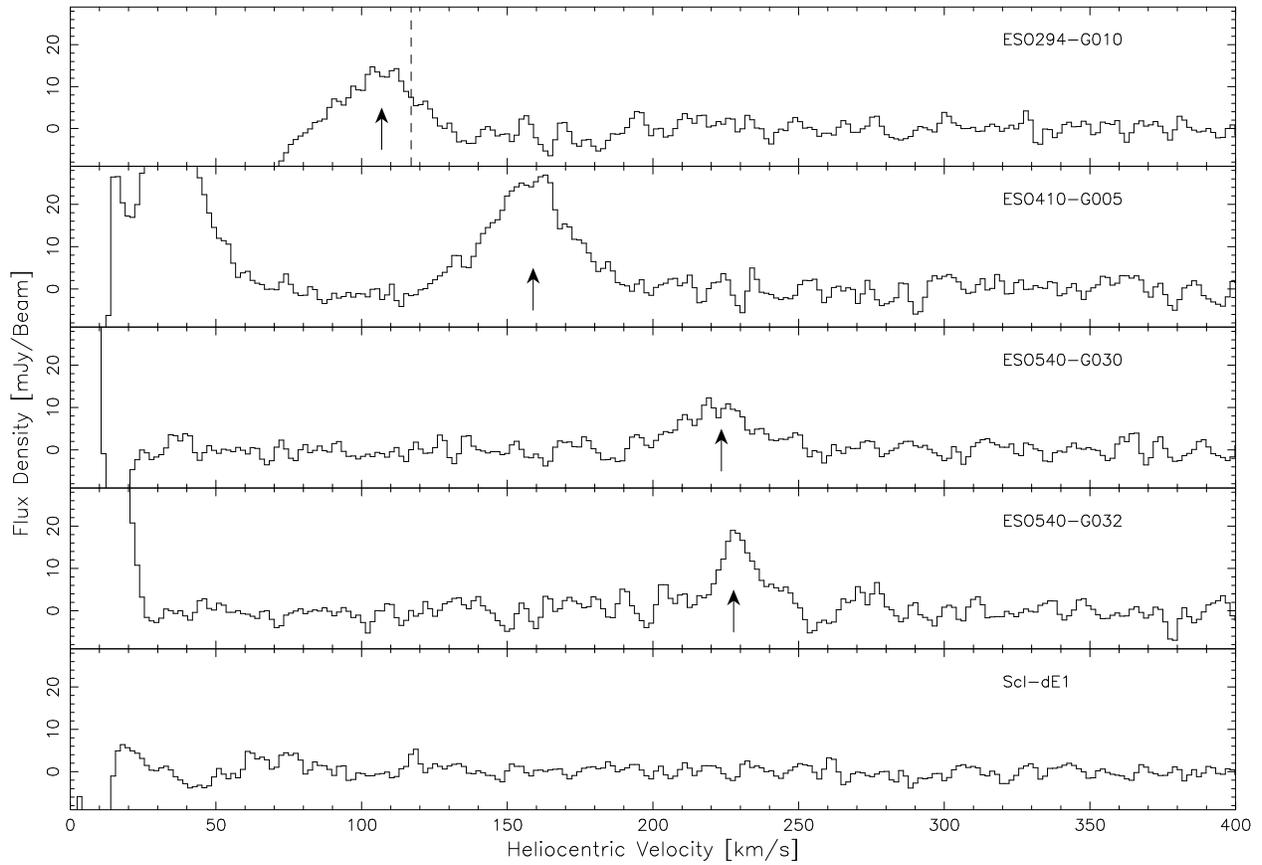}
\caption{Parkes \hi spectra of the five observed Sculptor dwarfs. The vertical
arrow indicates the central velocity of the 21\,cm emission line. In the top
panel the vertical dashed line shows the optical velocity of the dwarf. The
strong features in the 0--60\,\kms{} velocity range are from Galactic \hi
emission.}\label{spectra}
\end{center}
\end{figure}

The obtained spectra were reduced using the LIVEDATA data reduction pipeline.
The median of the Tukey smoothed bandpasses --- to remove the effect of radio
frequency interference --- was used for calibration. The gridding was performed
using GRIDZILLA, using the median of weighted values as estimator.  Both
LIVEDATA and GRIDZILLA are available in the AIPS++ software package.  To remove
residual baseline ripples, caused by a well known 5.8 MHz standing wave between
the focus cabin and the vertex of the Parkes telescope, the MBSPECT robust
fitting algorithm from the MIRIAD software package was used to fit (over the
complete spectral region) and remove  7$^{th}$ or 8$^{th}$ order polynomials.
The spectra of the five galaxies are shown in Fig.~\ref{spectra}. Various
moments (total flux, $\int S_{v}\,\textrm{d} v$, flux weighted mean velocity, $v_{\rm HI}$, and flux weighted velocity
dispersion, $\sigma_{\rm HI}$) are listed Table \ref{detections}. The errors listed were calculated
using a Monte-Carlo approach where normally distributed noise (the RMS was used
as the dispersion) was added to the spectra.

Data from the Australia Telescope Compact Array (ATCA) archive are available for
three of these objects: ESO294-G010, ESO540-G030, and ESO540-G032 (project
C705). These data were obtained in 1998 December in the 750D array and the
FULL\_8\_512 correlator configuration, corresponding to 8\,MHz bandwidth divided
in 512 channels. The source PKS1934-638 was used as flux calibrator and PKS0023-263
and PKS0022-423 were used for phase calibration. 

Taking advantage of our Parkes observations, we identified the channels in the
ATCA data that contain the 21\,cm line emission from the galaxies. Without the prior
knowledge of the \hi velocity range provided by the Parkes dataset, it would
have been impossible to detect the faint \hi signals near the target galaxies.

Using the MIRIAD data reduction package, the data were gridded using 'natural'
weighting to a channel spacing of 4~\kms{} and pixel size of
10$^{\prime\prime}$. The images were deconvolved using the CLEAN algorithm and
RESTORed to beam sizes of $69^{\prime\prime}\times52^{\prime\prime}$ for
ESO294-G010, $250^{\prime\prime}\times35^{\prime\prime}$ for ESO540-G030, and
$157^{\prime\prime}\times39^{\prime\prime}$ for ESO540-G032.  To
maximize the signal to noise ratio, we generated integrated intensity maps
using only those channels where the flux rose above 1$\sigma$ in the equivalent
Parkes spectra channel. The results are shown in Fig.~\ref{mom0} which also
shows optical images from the Digitized Sky Survey.

\section{Results and discussion}

\subsection{\hi association}

Galaxies in the Sculptor group have heliocentric velocities ranging from
$100<v<500$ \kms{} \citep{cote1997}. It was also previously established, by
direct distance measurements
\citep{jerjen1998,karachentsev2000,jerjen2001,karachentsev2003}, that the
galaxies we observed are genuine members of the Sculptor group.  Since our \hi
detections fall inside the velocity range mentioned above (our clouds have
velocities between 100 \kms{} and 250 \kms{}, Fig. \ref{spectra}) the assumption
that they are related to the targeted galaxies is reasonable.

However, we have no information on the distances to these \hi clouds and the
possibility of foreground 21\,cm emission with overlapping velocities cannot be
excluded a priori.  Because of the low velocity range and the projected position
of the Sculptor group on the sky, there are two potential contamination sources:
High Velocity Clouds (velocities between $-300$ \kms{} and $250$ \kms) and the
Magellanic Stream (between 0 \kms{} and 150 \kms{},
\citealt{putman2003,bruens2005}).

\begin{figure}
\plotone{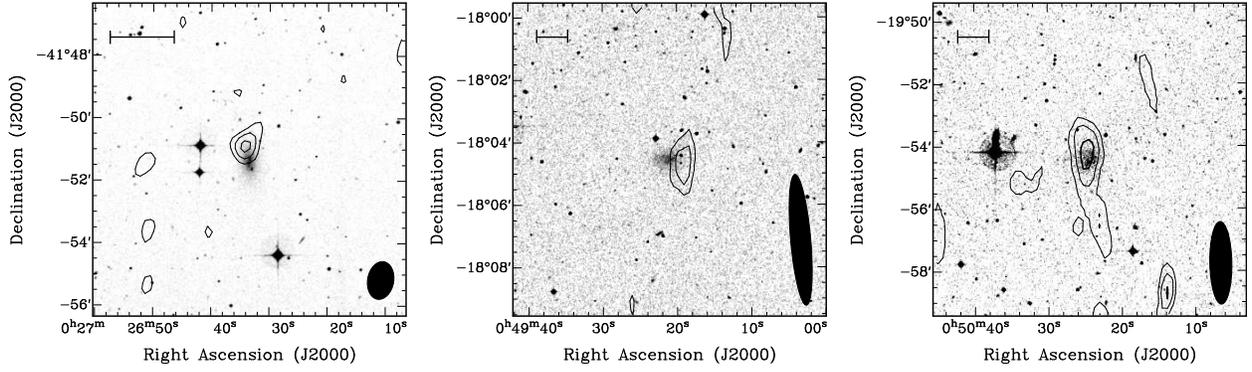}
\caption{Integrated intensity \hi maps (contours) overlayed on optical images
from the Digitized Sky Survey for
ESO294-G010 (left), ESO540-G030 (center), and ESO540-G032 (right). The 1$\sigma$
noise level corresponds to a column densities of 8.8$\times10^{18}$,
5.2$\times10^{18}$, and 6.2$\times10^{18}$ cm$^{-2}$ respectively. The contours
show the 2$\sigma$, 3$\sigma$ and 4$\sigma$ noise level. A scale of 1~kpc at the
distance of the object is
indicated in the top left corner of each frame whereas the filled ellipses on
the lower right corner are the beam sizes:
$69^{\prime\prime}\times52^{\prime\prime}$ for ESO294-G010,
$250^{\prime\prime}\times35^{\prime\prime}$ for ESO540-G030, and
$157^{\prime\prime}\times39^{\prime\prime}$ for ESO540-G032.}\label{mom0}
\end{figure}

High Velocity Clouds (HVCs) are Galactic \hi clouds with velocities that do not
fit simple Galactic rotation models.  We consider that HVC contamination is not
a source of confusion for the observed positions due to the following reasons.
First, there are no catalogued HVCs \citep{putman2002} within 30$^{\prime}$ of
any of our targets and that, in addition, there are no HVCs with velocities
consistent with those of our \hi detections within one degree.  Second, the
majority of Compact HVCs have sizes of the order of 30$^{\prime}$
\citep{braun2000,bruens2001,burton2001,deheij2002} although the most compact HVC
found to date has an extent of 4$^{\prime}$.4 \citep{bruens2004}, twice the size of
the present detections.  We, however, recognise that the \hi maps of Fig.
\ref{mom0} come from aperture synthesis observations, so that it is likely some
\hi flux remains undetected on the larger scales inaccessible with the array
configuration employed (a problem that \citealt{bruens2004} managed to avoid).  By
calculating the total \hi flux for the clouds from the ATCA data and comparing
these with the Parkes data, we estimate that the ATCA observations recover
$\sim40$\% of the flux for ESO294-G010, $\sim30$\% for ESO540-G030, and
$\sim80$\% for ESO540-G032. The clouds should therefore be slightly larger than
what is seen in Fig.~\ref{mom0}, but, nevertheless, they remain significantly
smaller than even the most extreme compact HVC. 

The Magellanic Stream remains a problem because of its large radial velocity
scatter near the Sculptor group \citep{bruens2005}. Further analysis would be
necessary if one wishes to disentangle the origins and establish the physical
location of each single cloud in that direction.  Nevertheless, there is
generally a good agreement between the position of the Sculptor dwarfs and the \hi
detections (Fig. \ref{mom0}). In the following, we will assume that the detected
\hi clouds are indeed associated with the Sculptor group galaxies.

\subsection{\hi properties}

For each object, the RMS of the spectra, the total flux, the flux weighted mean
velocity and velocity dispersion of the \hi are quoted in Table
\ref{detections}.  We also estimated the \hi masses $M_{\rm HI}$, expressed in
solar units, \msol, using the standard equation:
\begin{equation}
M_{\rm HI} = 2.356\times10^5~D^2~\int S_{v}\,\textrm{d} v
\end{equation}
\noindent
where $D$ is the distance in Mpc (in Table \ref{objects}), and $S_{v}$ is the
\hi flux density in Jansky.  The integration was computed up to where the signal fell
below the 1$\sigma$ noise level on either side of the emission peak. 

The derived total \hi masses for ESO294-G010, ESO410-G005, ESO540-G030 and
ESO540-G032 are $M_{\rm HI}$~=~(3.0$\pm$0.3)$\times$10$^5$~\msol{} (the peak
flux is 8$\sigma$ above noise level), 
$M_{\rm HI}$~=~(7.3$\pm$1.5)$\times$10$^5$~\msol{} (12$\sigma$), 
$M_{\rm HI}$~=~(8.9$\pm$1.9)$\times$10$^5$~\msol{} (7$\sigma$) and 
$M_{\rm HI}$~=~(9.5$\pm$1.6)$\times$10$^5$~\msol{} (9$\sigma$) respectively. The
undetected galaxy, Scl-dE1, has an upper limit $M_{\rm
HI}~<~1.0\times$10$^5$~\msol{} (3$\sigma$ limit for an assumed 10 \kms{}
velocity dispersion at 4.21 Mpc). This is comparable to Tucana which has no \hi
detected within its optical boundary \citep[$M_{\rm HI}$/$L_{\rm B}$ $<$ 0.03
$M_{\odot}/L_{\odot}$,][]{oosterloo1996}.  The \hi masses for the other four
galaxies are comparable to that of similar Local Group mixed and early type
dwarfs where \hi has been detected: Sculptor dSph has 2.3$\times10^5$ \msol{}
\citep[$M_{\rm HI}$/$L_{\rm B}$ = 0.2 $M_{\odot}/L_{\odot}$,][]{bouchard2003},
LGS3 has 4.2$\times10^5$ \msol{} \citep[$M_{\rm HI}$/$L_{\rm B}$ = 0.3
$M_{\odot}/L_{\odot}$,][]{young1997}, Phoenix has 1.9$\times10^5$ \msol{}
\citep[$M_{\rm HI}$/$L_{\rm B}$ = 0.2 $M_{\odot}/L_{\odot}$,][]{st-germain1999},
and Antlia exhibits 6.8$\times10^5$ \msol{} \citep[$M_{\rm HI}$/$L_{\rm B}$ =
0.4 $M_{\odot}/L_{\odot}$,][]{barnes2001}.


\citet{buyle2005} argued that the low values of $M_{\rm HI}$/$L_{\rm B}$
typically found in dEs may be a direct result from a near complete gas depletion
due to enhanced star formation efficiency. They were using the results of an
analytical chemical evolution model \citep{pagel1998} and stellar
mass-to-light ratios ($M_{\star}$/$L_{\rm B}$) from a simple stellar population
model \citep[SSP][]{vazdekis1996}. In this approach all the gas that has been
ejected from stars (stellar winds, supernova explosions, planetary nebulae,
etc.) is effectively expelled from the galaxy, but leaving part of the metals to
enrich the chemical composition of subsequent gas inflow and star formation. The
resulting state is a metal-enhanced (unless the metals also escape) but
inevitably gas deficient galaxy. 

Using the SSP model from \citet{anders2003} at a metallicity (Z) of 0.02
Z$_{\odot}$ and by
assuming a constant star formation rate of 0.01 $M_{\odot}$ per yr over the
entire lifetime of the galaxy, we calculated that the ISM return from normal
stellar evolution should give a mass-to-light ratio of the order of $M_{\rm
ISM}$/$L_{\rm B}$ = 0.25 $M_{\odot}/L_{\odot}$ after 10 Gyr.  Our objects have
$M_{\rm HI}$/$L_{\rm B}$ between 0.08 $M_{\odot}/L_{\odot}$ and 0.19
$M_{\odot}/L_{\odot}$, consistent with this simple case of passive, undisturbed
evolution. Most of the expelled gas should therefore be retained in the galactic
potential well and cooled to the form of H{\sc i}. This, however, implies that
the pristine gas used to form stars in the galaxy, has not build-up any
reservoirs and is instantly used up by star formation. We must therefore
conclude that the \hi content in our galaxies is consistent with both the
enhanced efficiency of star formation and gas buildup from stellar evolution. In
the case of Scl-dE1, like many Local Group early-type dwarfs \citep[as noted by
][]{grebel2003} and even Galactic globular clusters, some factors may have
prevented the cooling process and/or the accumulation of \hi in the system. 

\begin{figure}
\begin{center}
\plotone{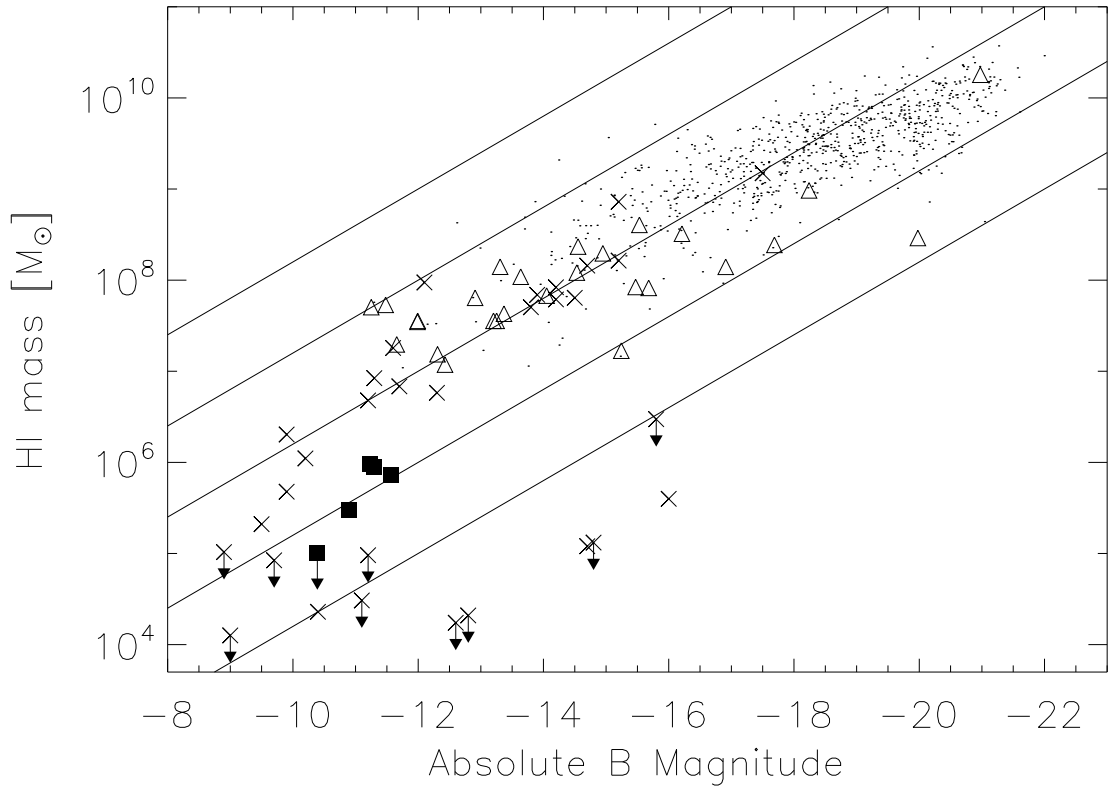}
\caption{The \hi mass versus absolute B band magnitude of a selection of galaxies. The filled
squares are the five galaxies from this paper, the triangles are the galaxies of
the Sculptor and Centaurus A group where data is available, the crosses show the
location of Local Group galaxies \citep{mateo1998} and the points denote data
taken from the HIPASS Bright Galaxy Catalog \citep{koribalski2004}. The five
diagonal lines show constant $M_{\rm HI}$/$L_{\rm B}$ of, from top to bottom, 100, 10, 1,
0.1 and 0.01 $M_{\odot}/L_{\odot}$.}\label{mtol}
\end{center}
\end{figure}

Fig. \ref{mtol} shows the relation between \hi mass and absolute magnitude for
galaxies from the HIPASS Bright Galaxy Catalog \citep{koribalski2004}, from the
Centaurus\,A and Sculptor Group \citep{cote1997} and from the Local Group
\citep[][ and references therein]{mateo1998}.  Our results agree with that of
\citet{warren2004}, who noted that dwarf galaxies have a larger spread in
$M_{\rm HI}$/$L_{\rm B}$ than brighter galaxies.  There is indeed evidence that
the spread is one order a magnitude larger in the dwarf regime compared to more
luminous galaxies.

\subsection{\hi distribution}

The \hi maps in Fig. \ref{mom0} show, in two out of three cases, an apparent
offset between the \hi gas and stellar components of the galaxies. Because of
the low signal to noise ratio in our aperture synthesis images we adopted a
conservative error on the accurate position of the \hi emission peak of a third of the
beam size. These offsets are $\sim$35$^{\prime\prime}$ (290 pc projected
distance or $\sim0.9$ times its Holmberg radius, $R_{\rm Ho}$,
\citealt{jerjen2000}) for ESO294-G010 --- greater than the
$\sim20^{\prime\prime}$ accuracy --- and $\sim$30$^{\prime\prime}$ (460 pc or
$\sim0.85\,R_{\rm Ho}$) for ESO540-G030 --- greater than the
$\sim10^{\prime\prime}$ accuracy. In the third case, ESO540-G032, the apparent
offset is smaller than the uncertainty of the \hi position and thus consistent
with being centered.

An offset between the \hi and optical was previously detected in the Local Group
galaxy Phoenix. Not only is the \hi in Phoenix found 5$^{\prime}$ or 650 pc from
the optical center of the galaxy \citep{st-germain1999} but \citet{gallart2001}
also found that the gas is kinematically separated from the optical by 29
\kms{}. These authors have also conducted semi-analytical and numerical
analyses to demonstrate that the \hi offset is either the consequence of the
environmental conditions, i.e. ram pressure by the intergalactic medium Phoenix
travels through or the result of an internal process i.e. several supernova
explosions in the dwarf.  The authors favored the former explanation --- without
discarding the latter --- because it produced a smooth, anisotropic structure
similar to that observed. 

In the low density environment represented by the Sculptor Group, most galaxies
can be considered as essentially isolated objects, and, in that respect, should
be under similar conditions as Phoenix (which lies at 450 kpc from the Milky
Way).  More precisely, the closest neighbour to both ESO540-G030 and ESO540-G032
is the faint IB(s)m galaxy DDO6, at a relative 3-dimensional distance of 180 kpc
and 100 kpc, respectively, while the brighter spiral NGC247 is around 700 kpc
behind the pair.  The closest neighbour of ESO294-G010 is the spiral NGC55 at a
3-dimensional distance of 160 kpc \citep[radial distances from
][]{karachentsev2004}.  Although these distances are significantly smaller than
that of Phoenix to the Milky Way, the larger counterparts are of much lower
mass. We can therefore draw similar conclusions as \citet{gallart2001} to
explain the \hi and optical geometry of the systems.  The possibility of tidal
stripping seems unlikely in this case as the gas would not be concentrated on
one side of the optical but stretched on both sides.  Furthermore, with the
large distances separating the dwarfs to their neighbours, the tidal forces
should be minimal.

It is worth noting that, in the case of ESO294-G010, \citet{jerjen1998}
reported the existence of an H{\sc ii} region $\sim18^{\prime\prime}$ south of
its optical center while the \hi is on the north side. This is the opposite
situation as was observed in Phoenix were the \hi emission is on the same side as
the galaxy's youngest stars \citep{martinez1999}. This might suggest that in
ESO294-G010, the supernova explosion/stellar wind scenario might play a bigger
role than in the case of Phoenix. Detailed stellar population studies of these
Sculptor dwarfs are needed to explore the various possibilities. 

The only Sculptor galaxy where no \hi is detected to date, Scl-dE1, is also one
of the only two dEs of the group. Yet, contrary to the prediction of the
morphology-density relation, this object is located far from any other group
members. Its closest more massive neighbour, the starburst spiral NGC253, is
spatially separated by $\sim480$ kpc (5.4$^{\circ}$ projected angular separation
or $\sim400$ kpc tangential distance, and 270 kpc radial distance). We note that
the spiral galaxy NGC45 is the closest in projection but its distance of 5.9 Mpc
(\citealt{tully1988}) places it 2\,Mpc, to the far end of the group.  This
situation of Scl-dE1 is reminiscent of the case of the isolated Local Group
early-type dwarf Tucana (880 kpc away from the Milky Way) since both objects
also have predominantly old stellar populations, or the Cetus dwarf (755 kpc from
the Milky Way) which has some young stars \citep{mcconnachie2005} but no H{\sc
i}.  Apart from the possibility of Scl-dE1, Tucana, and Cetus being on highly
eccentric orbits  that would lead these galaxies much closer to their
neighbouring galaxies at perigalacticon, there is only a weak case for an
environmentally driven gas removal process or ionisation and thus strong
internal mechanisms (e.g. Supernovae driven winds or radiative feedback, see
\citealt{dekel2003}) must be responsible for the properties of these
objects.

\section{On the Sculptor group and the morphology-density relation}
The Sculptor group is known to be a loose aggregation of several luminous
galaxies with satellites and late-type dwarfs populating the space between rather than a
virialized system \citep{jerjen1998}. Fig.  \ref{distribution} shows the large
extent of the galaxy distribution on the sky  of $\sim25^{\circ}$ or over 1\,Mpc
across, at an assumed mean distance of 2.5\,Mpc. Group members exhibit even a
wider dispersion in their line of sight distances
\citep{jerjen1998,karachentsev2003}.  \citet{cote1997} initially derived a
dynamical crossing time of 3.2$\times10^9$ years assuming a virialized system and using
projected distances to the tentative group center. Because of the ``cigar-like
distribution'' \citep{jerjen1998} of the group the assumption of an isotropic
system does not stand and this value can only be a lower limit whereas the true
value must be closer to a Hubble time. This makes the Sculptor group a
prototypical low density environment where interaction between galaxies are
expected to be minimal and external gas stripping largely ineffective.

Indeed, all but one galaxy of the Sculptor group have substantial amounts of
H{\sc i} which is either primordial or the result from stellar feedback. In all
cases, the environment of these galaxies has not managed to either remove that
gas or ionise it. This can be viewed as a confirmation of the overall validity
of the morphology-density relation in a low density environment. There is,
however, one caveat: contrary to the standard view of the morphology-density
relation and the environmentally driven late-type to early-type evolution
scenario, the low mass end of the galaxy spectrum in the loosely-bound Sculptor
group is not solely populated by late-type dwarfs. In fact, our results support
the \citet{grebel2003} contention that: {\it ``Transition-type dwarfs are dSphs
that kept their interstellar medium and therefore should replace dSphs in
isolated locations where stripping is ineffective.''} This is further reinforced
by the fact that our five target galaxies are of very low $M_{\rm HI}/L_{\rm B}$
(see Table \ref{detections} and Fig.  \ref{mtol}). 

In line with the morphology-density relation and because of the presence of both
\hi gas and a centrally concentrated population of young blue stars
\citep{karachentsev2000} in ESO410-G005, we propose a reclassification of
ESO410-G005 (initially labelled dSph, \citealt{karachentsev2000}) to dSph/dIrr.
We believe that this reclassification would better reflect the nature of this
low mass dwarf. Similarly, since NGC59 contains both \hi \citep{cote1997} and
H{\sc ii} regions \citep{skillman2003a}, it is not a genuine early-type dwarf
and should be classified as dS0 pec.  These changes give weight to the
empirical relation between morphology and ISM content in the Sculptor group,
i.e. a slightly irregular morphology caused by the presence of young stars seem
to indeed indicate an underlying presence of ISM.  Consequently, the Sculptor
group has only one single early-type dwarf system: Scl-dE1, without H{\sc i} or
any young stars \citep{karachentsev2003}.

This, however, highlights a second caveat in the morphology-density relation for
low density environments. While the relation describes well the general trend of
gas-rich dwarf irregular galaxies being relatively isolated stellar systems and
early-type dwarfs the satellites of more luminous galaxies, it provides no
explanation for the existence of an isolated gas deficient early-type dwarf
galaxy like Scl-dE1. This object joins a list of similar systems including
Tucana and Cetus that do not fit the global picture. Clearly these systems hold
important clues to what extent internal gas expulsion mechanisms (supernova
explosions or stellar winds) govern the passive evolution of dwarf galaxies in
the field.

\section{Conclusions}
In this paper we have presented the results of \hi observations toward the five
lowest luminosity dwarf galaxies of the Scultor group. The main results are:
\begin{enumerate}
\item All but one galaxy of the Sculptor group have substantial amounts of \hi
($M_{\rm HI}\, >\,3\times 10^5$ \msol{}). Four new \hi detections have been made
with $3\times10^5\,$\msol$<\,M_{\rm HI}\,<\,10^6\,$\msol. The only undetected
galaxy, Scl-dE1, must have $M_{\rm HI}\, <\,10^5$ \msol{}. All Sculptor galaxies
except Scl-dE1 have $M_{\rm HI}$/$L_{\rm B}\, >\,0.08$ $M_{\odot}/L_{\odot}$.
\item The \hi masses for ESO294-G010, ESO410-G005, ESO540-G030, and ESO540-G032
are all consistent with passive, undisturbed evolution where all the gas from
stellar winds has condensed in the form of H{\sc i}.
\item ESO410-G005, formerly classified as a dSph galaxy, better fits the
dSph/dIrr category while NGC59 is a dS0 pec. There is only one genuine
early-type dwarf galaxy in the Sculptor group: Scl-dE1.
\item At least two of the mixed-morphology galaxies, ESO294-G010 and
ESO540-G030, have a significant offset between their optical and \hi components.
These offsets are likely to be caused by winds from star forming regions.
\item The morphology-density relation seems to be valid for low-density
environments where mixed-type galaxies, and not late-types, replace the
early-type systems when in isolation.
\item Scl-dE1 joins Tucana and Cetus as an isolated low mass gas-deficient
early-type dwarf. These objects require internal mechanisms
to explain their evolutionary states.
\end{enumerate}

\acknowledgments
We would like to thank Sylvie Beaulieu, Lister Staveley-Smith and B\"arbel
Koribalski for insightful discussions, as well as the anonymous referee for
diligent and valuable comments. This research has been supported by the
Australian Research Council through Discovery Project Grant DP0343156. The
Digitized Sky Surveys were produced at the Space Telescope Science Institute
under U.S. Government grant NAG W-2166.

\clearpage

\clearpage

\clearpage

\clearpage

\clearpage

\begin{table}[tbp]
\begin{center}
\caption{Position and optical parameters of the sample}
\label{objects}
\begin{tabular}{l c c c c c c c}
\hline
\hline
Galaxy		& Type	& RA		& Dec		& D	& $m_{\rm B}$	 & $L_{\rm B}$		& $v_{\rm opt}$\\
		&	& (J2000)	& (J2000)	& (Mpc)	&    	 & ($10^5$ L$_{B,\odot}$) & (km s$^{-1}$)	\\
\hline
ESO294-G010	& dS0/dIrr	& 00 26 33.4	& -41 51 19	& 1.92$\pm$0.10\tablenotemark{a}	& 15.53$\pm$0.04\tablenotemark{b} & 36.9$\pm$4.1	& 117$\pm$0.8\\
ESO410-G005	& dSph/dIrr	& 00 15 31.4	& -32 10 47	& 1.92$\pm$0.19\tablenotemark{c}	& 15.12$\pm$0.1\tablenotemark{c} & 55.3$\pm$12.2  	& \nodata \\
ESO540-G030	& dE/dIrr	& 00 49 21.1	& -18 04 34	& 3.40$\pm$0.34\tablenotemark{d}	& 16.37$\pm$0.07\tablenotemark{b} & 56.9$\pm$12.1  	& \nodata \\
ESO540-G032	& dE/dIrr	& 00 50 24.5	& -19 54 23	& 3.42$\pm$0.27\tablenotemark{d}	& 16.44$\pm$0.08\tablenotemark{b} & 53.5$\pm$9.4 	& \nodata \\
Scl-dE1		& dE	& 00 23 51.7	& -24 42 18	& 4.21$\pm$0.43\tablenotemark{d}	& 17.73$\pm$0.018\tablenotemark{b} & 24.0$\pm$6.4 	& \nodata \\
\hline
\end{tabular}
\tablenotetext{a}{\citet{karachentsev2002}}
\tablenotetext{b}{\citet{jerjen2000}}
\tablenotetext{c}{\citet{karachentsev2000}}
\tablenotetext{d}{\citet{karachentsev2003}}
\end{center}
\end{table}

\begin{table}[tbp]
\begin{center}
\caption{Parkes \hi detections}
\label{detections}
\begin{tabular}{l c c c c c c c c}
\hline
\hline
Galaxy		& RMS		&  $\int S_{v}\,\textrm{d} v$ & $v_{\rm HI}$	& $\sigma_{\rm HI}$	& $M_{\rm HI}$	& $M_{\rm HI}$/$L_{\rm B}$ \\
		& (mJy)		&  (mJy km s$^{-1}$) & (km s$^{-1}$)	& (km s$^{-1}$)		& ($10^5$ M$_{\odot}$)	&	(M$_{\odot}$/L$_{\odot}$) 	\\
\hline
ESO294-G010	& 1.8		&  342$\pm$17	& 106.9$\pm$0.8	& 9.7$\pm$0.9	& 3.0$\pm$0.3	& 0.08$\pm$0.01\\
ESO410-G005	& 2.3		&  835$\pm$34	& 158.9$\pm$1.9	& 14.2$\pm$0.6	& 7.3$\pm$1.5	& 0.13$\pm$0.04\\
ESO540-G030	& 1.8		&  327$\pm$25	& 223.5$\pm$2.7	& 11.0$\pm$1.2	& 8.9$\pm$1.9	& 0.16$\pm$0.05\\
ESO540-G032	& 2.1		&  346$\pm$19	& 227.7$\pm$0.9	& 7.4$\pm$1.2	& 9.5$\pm$1.6	& 0.18$\pm$0.04\\
Scl-dE1		& 1.2		&  $<$ 22.5	& \nodata	& (10)		& $<$1.0	& $<$0.04 	\\
\hline
\end{tabular}
\end{center}
\end{table}

\end{document}